\documentclass[elsart12,epsf,eqsecnum,graphics,cite]{revtex4}
\def\beq{\begin{equation}}

\def\eeq{\end{equation}}

\usepackage{epsfig,epsf}
\usepackage{graphicx}
\usepackage{grffile}
\usepackage{amsmath,amsfonts,amssymb,amsthm,nccmath,latexsym,mathtools} \usepackage[mathscr]{euscript}

\newcommand{\be}{\begin{equation}}
\newcommand{\ee}{\end{equation}}
\usepackage{hyperref}

\begin{document}


\voffset1.5cm

\title{Conformal Gravity Redux: Ghost-turned-Tachyon.}
\author{Ibrahim Burak Ilhan$^{1}$ and Alex Kovner$^{1}$}

\affiliation{
$^1$ Physics Department, University of Connecticut, 2152 Hillside
Road, Storrs, CT 06269-3046, USA}

\begin{abstract}
We analyze conformal gravity in translationally invariant approximation, where the metric is taken to depend on time but not on spatial coordinates. We find that the field mode which in perturbation theory has a ghostlike kinetic term, turns into a tachyon when nonlinear interaction is accounted for. The kinetic term and potential for this mode have opposite signs. Solutions of nonlinear classical equations of motion develop a singularity in finite time determined by the initial conditions. 
\end{abstract}
\maketitle
\section{Introduction} 
Recent years have seen a surge of interest in generalizations of the theory of general relativity. One strong motivation for this is the discovery of cosmic acceleration\cite{acceleration} and the associated need for a non-vanishing cosmological constant, which has no natural explanation within general relativity.  One can hope that modifying gravitational interactions at large distance scales might bring a natural understanding of this problem. 

Another problem where modifying gravity can potentially bring dividends is dark matter. Dark matter has not been observed directly, although within the standard cosmological model it is necessary to account for the energy balance of the universe, as well as explaining rotation curves of galaxies.

Conformal gravity is an example of a modified theory of gravity which is potentially interesting in both these contexts\cite{mannheim}. Simple two parameter fits based on conformal gravity describe all available galactic rotation curves very well\cite{mannheim1}. Arguments for naturalness of the cosmological repulsion in conformal gravity have also been presented\cite{mannheim2}. An important aspect of conformal gravity that singles it our from other higher derivative extensions of GR is that it is renormalizable by power counting in the ultraviolet \cite{stelle}, and on this basis has been considered as a candidate for a consistent quantum theory of gravity.

It is however not clear whether conformal gravity is consistent. The problem, like with many higher derivative theories, is that in perturbation theory it has ghost modes - the modes whose kinetic energy is negative. As long as interactions between the field modes are neglected, the wrong sign of kinetic energy is not a problem as such. Since in a free approximation any field theory has infinite number of conserved quantities, the classical motion of such a system is bounded. All the oscillators simply oscillate independently of each other, and the sign of the energy for each one is a matter of convention. 

However, once interactions between the modes are turned on, one generally expects that the classical motion  becomes ergodic, and samples all available phase space. If the total energy is not bounded from below, this is expected to lead to classical instability with positive and negative contributions to energy growing without bound.
Sometimes the ghosts are said to violate unitarity of a quantum theory. As explained, for example in \cite{pais} this is simply another way of stating the same problem. In such a quantum system time evolution evolves a normalizable quantum state into a state which has support only for ``infinite'' values of the field, thereby ``violating unitarity''. A classical theory with such behavior cannot yield a consistent quantum field theory upon quantization. 

The problem of ghosts, or unitarity afflicts many extensions of gravity\cite{ghost}. For example massive gravity has perturbatively a ghost mode and much effort has been spent to understand whether this ghost can be consistently decoupled\cite{gabadadze}. It has been however convincingly argued recently that one does not need to decouple the ghost, since nonperturbatively the theory cures itself and the full nonlinear Hamiltonian of spontaneously broken gravity is bounded from below\cite{zurab}.

In fact in simple quantum mechanical systems presence of ghosts does not immediately signals instability, even if the theory is interacting. Some consistent simple models with interacting ghost and normal modes have been discussed in \cite{pais},\cite{smilga},\cite{pavic}. In a quantum field theory such stability must be much harder to achieve due to many excitation channels available\cite{jeon}. Nevertheless it is an interesting open question, whether the ghost modes in conformal gravity do indeed render the full interacting  theory unstable, or perhaps the theory is consistent ``as is''\cite{footnote1}. In fact it has been shown that the number of local conserved quantities in conformal gravity is equal to the number of perturbative ghost modes \cite{exam}. This can give hope that the dynamics is constrained enough and not ergodic to an extent that instabilities do not appear even in the interacting theory. 

Complete analysis of an interacting theory of gravity is a very complicated proposition. Our aim in this paper is much more modest. We ask if the theory has instabilities when the number of degrees of freedom is restricted to translationally invariant modes. The requirement of translational invariance is very severe and reduces the field theory to a theory of a finite albeit relatively large number of classical degrees of freedom. We derive the Hamiltonian for this system and study classical behavior of its solutions. Our result is somewhat unexpected. We find that the theory is unstable on the classical level. The instability is of a somewhat different nature than what we may have expected from the previous argument. It is not due to transfer of large amount of energy from ghost modes to normal modes. Instead the nonlinearity of the interaction induces a potential for the ghost modes which is positive. Thus the ghost becomes also a tachyon - it's kinetic term is negative, while its potential is positive. Thus the ghost sector becomes unstable by itself. We find simple classical solutions for which normal modes are vanishing, and ghost modes diverge within a finite amount of time, set by the initial conditions.

The plan of this paper is the following. In Section 2 we derive the Hamiltonian of conformal gravity in the translationally invariant approximation and discuss the symmetries of the reduced model that follow from the gauge symmetries of the full theory. In Section 3 we transform the model into a set of simple degrees of freedom, and exhibit some classical solutions which exhibit the properties we alluded to earlier. Finally in Section 4 we discuss our results.

\section{The Hamiltonian of the Reduced Theory}
Conformal gravity is defined by the action
\be S= - \int d^4x \sqrt{-g} (3 R_{\mu\nu}R^{\mu\nu}-R^2)\ee
with the usual definitions of the Riemann and Ricci tensors ${R^\rho}_{\mu\sigma\lambda} = - \partial_\sigma {\Gamma^\rho}_{\mu\lambda}+...$ and $R_{\mu\lambda} = {R^\sigma}_{\mu\sigma\lambda}$. We use the metric convention $ (+,-,-,-)$. Since our interest is in the classical theory, we set the dimensionless coupling constant to unity, as its value does not affect solutions of equation of motion.

We treat this Lagrangian as a Lagrangian of an ordinary field theory. We will derive the Hamiltonian which generates classical time evolution by Legandre transforming it rather than using the ADM procedure\cite{adm}. Since the Lagrangian possesses gauge invariance, this is of course a constrained system, and constraints have to be properly taken into account.
The Lagrangian, as is well known is gauge invariant under the general linear transformation
\be g_{\rho\sigma} (x)\rightarrow g_{\rho\sigma}' (x') = g_{\mu\nu}(x) \frac{\partial x^\mu}{\partial {x^\rho}'}\frac{\ x^\nu}{\partial {x^\sigma}'}\ee
and, in addition the local conformal transformation:
\be g_{\mu\nu}(x)\rightarrow \tilde{g}_{\mu\nu}(x)=\Omega^2(x)g_{\mu\nu}(x)\ee
We choose to impose a simple gauge fixing condition:
\be \label{gaugef}g_{00}=1, \ \ \ \ g_{i0}=0.\ee
This gauge condition does not fix one combination of conformal and general linear transformations (see Appendix), and we will deal with this remaining gauge
symmetry later.

We truncate the theory by taking the metric to be space independent $g_{\mu\nu}=g_{\mu\nu}(t)$.
The non vanishing components of the Christoffel symbol and Ricci tensor, in the gauge eq.(\ref{gaugef}) for metric that does not depend on spatial coordinates, are:

\be {\Gamma^0}_{ij}= -\frac{1}{2}\partial g_{ij}, {\Gamma^i}_{0j}=\frac{1}{2}g^{ik}\partial g_{jk}\ee

\be R_{00}=\partial {\Gamma^i}_{0i} + {\Gamma^i}_{j0}{\Gamma^j}_{i0}= \frac{1}{2}\partial(g^{ij}\partial g_{ij})+ \frac{1}{4}g^{ik}\partial g_{kj}g^{jm}\partial g_{mi}=\frac{1}{2}\partial \alpha - \frac{1}{4}\beta\ee
\be R_{ij} = -(\partial {\Gamma^0}_{ij} + {\Gamma^k}_{k0}{\Gamma^0}_{ij}) + ({\Gamma^0}_{kj}{\Gamma^k}_{0i}+ {\Gamma^k}_{0j}{\Gamma^0}_{ki})= \frac{1}{2}\partial^2 g_{ij} + \frac{1}{4}\alpha \partial g_{ij} - \frac{1}{2} {\alpha^k}_j\partial g_{ki}\ee
\be R=\frac{1}{4}\partial \alpha - \frac{1}{4}\beta + \frac{1}{4}\alpha^2\ee

where, we have defined:
\be {\alpha^i}_j = g^{ik}\partial g_{kj}; \ \ \ \ \alpha = {\alpha^i}_i\ee
\be {\beta^i}_j = \partial g^{ik}\partial g_{kj};\ \ \ \  \beta = {\beta^i}_i\ee

The action can be written as:
\begin{eqnarray} S= &-& \int dt \sqrt{-g} \left[(3 \left((\frac{1}{2}\partial \alpha - \frac{1}{4}\beta)^2 + (\frac{1}{2}\partial {\alpha^a}_j + \frac{1}{4}\alpha {\alpha^a}_j)(\frac{1}{2}\partial {\alpha^j}_a + \frac{1}{4}\alpha {\alpha^j}_a)\right)-\left((\frac{1}{2}\partial\alpha -\frac{1}{4}\beta) + (\frac{1}{2}\partial\alpha + \frac{1}{4}\alpha^2)\right)^2\right] \nonumber\\
 &=&- \int dt \sqrt{-g}\left[-\frac{1}{2}(\beta + \alpha^2)(\frac{1}{2}\partial \alpha - \frac{1}{4}\beta) + 3(\frac{1}{4} \partial {{\tilde{\alpha}^a}}_b \partial {{\tilde{\alpha}^b}}_a+ \frac{1}{4}\alpha {{\tilde{\alpha}^a}}_b \partial {{\tilde{\alpha}^b}}_a + \frac{1}{16}\alpha^2 {{\tilde{\alpha}^a}}_b{{\tilde{\alpha}^b}}_a)\right]\end{eqnarray}

Where, ${\tilde{\alpha}^a}_b$ is the traceless part of ${\alpha^a}_b$
\be {\tilde{\alpha}^a}_b={\alpha^a}_b+\frac{1}{3}\alpha g^a_b\ee




After some simple manipulations, involving integration by parts, this can be written as
\be \label{action}S= - \int dt \sqrt{-g}\left[\frac{3}{4}\partial {{\tilde{\alpha}^a}}_b \partial {{\tilde{\alpha}^b}}_a - \frac{1}{8}\partial \alpha {\rm tr} (\tilde\alpha^2)-\frac{1}{24}\alpha^2{\rm tr} (\tilde\alpha^2)  +\frac{1}{8}\left[{\rm tr} (\tilde\alpha^2)\right]^2\right]\ee
Or using the identity
\be \partial[\sqrt{-g} \left[{\rm tr} (\tilde\alpha^2)\right]] = \frac{1}{2}\sqrt{-g} \alpha^2   \left[{\rm tr} (\tilde\alpha^2)\right] + \sqrt{-g}\partial \alpha  {{\tilde{\alpha}}^a}_b \partial  {{\tilde{\alpha}}^b}_a + 2 \sqrt{-g}\alpha  {{\tilde{\alpha}}^a}_b \partial  {{\tilde{\alpha}}^b}_a\ee
and integrating by parts
\be  S= - \int dt \sqrt{-g} \left[3\partial  {{\tilde{\alpha}}^a}_b \partial  {{\tilde{\alpha}}^b}_a + \alpha  {{\tilde{\alpha}}^a}_b \partial  {{\tilde{\alpha}}^b}_a + \frac{1}{2} {\rm tr} (\tilde\alpha^2)\left({\rm tr} (\tilde\alpha^2) + \frac{\alpha^2}{6}\right)\right]\ee
The latter form is more convenient for applications since it makes it obvious that no time derivatives of $\alpha$ appear in the action.
 
Since we imposed gauge conditions in the action, we must in principle separately keep track of constraints that would be generated by variation of the action with respect to $g_{00}$ and $g_{i0}$. However in our reduced theory this turns out not to be necessary. The variation of the action with respect to $g_{\mu 0}$ results in the equations
\be \label{cons}B_{\mu0}=0\ee  where $B_{\mu\nu}$ is the so called Bach tensor:
\be  B_{\mu\nu} \equiv \nabla^\alpha \nabla^\beta C_{\mu\alpha\nu\beta} - \frac{1}{2}R^{\alpha\beta}C_{\mu\alpha\nu\beta} =0.\ee
Here $C_{\mu\alpha\nu\beta}$ is the conformal tensor - the traceless part of the Riemann tensor:
\be C_{\mu\nu\alpha\beta} = R_{\mu\nu\alpha\beta}-(g_{\mu[\alpha}R_{\beta]\nu}-g_{\nu[\alpha}R_{\beta]\mu}) + \frac{1}{3}Rg_{\mu[\alpha}g_{\beta]\nu}.\ee

However, in the gauge $g_{i0}$ in the reduced theory (no $x_i$ dependence) it is obvious that $B_{i0}=0$ identically. The Bach tensor is by definition traceless, thus identically
\be B_{00}=g^{ij}B_{ij}\ee
Therefore $B_{00}$ vanishes automatically when the spatial components vanish. These are required to vanish by equations of motion that follow from the action eq.(\ref{action}). Thus in the translationally invariant approximation, constraints eq.(\ref{cons}) do not add any new information, and we can forget about their existence.

\subsection{The Hamiltonian.}
Our aim now is to derive the Hamiltonian for the system described by the action eq.(\ref{action}). 
Since the fields $\alpha$  are related to the time derivative of $g_{ij}$, we introduce this relation into the action with the help of the Lagrange multiplier
\be  S= - \int dt \sqrt{-g} \left[3\partial  {{\tilde{\alpha}}^a}_b \partial  {{\tilde{\alpha}}^b}_a + \alpha  {{\tilde{\alpha}}^a}_b \partial  {{\tilde{\alpha}}^b}_a + \frac{1}{2} {\rm tr} (\tilde\alpha^2)\left( {\rm tr} (\tilde\alpha^2) + \frac{\alpha^2}{6}\right) - {\lambda^a}_b({\alpha^b}_a - g^{bc}\partial g_{ca})\right]\ee
The conjugate momenta are:
\be p^{ij} = \frac{\partial L}{\partial (\partial g_{ij})}=-\frac{\sqrt{-g}}{2} [{\lambda^{i}_b}g^{jb}+{\lambda^{j}_b}g^{ib}]\label{lambda}\ee
\be {\beta^i}_j=\frac{\partial L}{\partial(\partial {\tilde{\alpha}^j}_i)}= -\sqrt{-g} (6\partial {\tilde{\alpha}^i}_j + \alpha {\tilde{\alpha}^i}_j ) \label{bbeta} \ee

and
\be p_\alpha=p_\lambda=0
\ee

To find the Hamiltonian, we take the Legendre transform of the action and use eq.(\ref{lambda}) to express $\lambda^a_b$ in terms of the momenta $p^{ij}$. The resulting Hamiltonian is
\be H= p^{ij}\partial g_{ij} - L = -\frac{1}{\sqrt{-g}}\frac{1}{12} {\beta^i}_j {\beta^j}_i + \frac{1}{6}\alpha {\tilde{\alpha}^i}_j {\beta^j}_i + \frac{1}{2}\sqrt{-g} \left[{\rm tr} (\tilde\alpha^2)\right]^2- {\alpha^b}_a p^{an}g_{nb} \label{hamil}\ee
The is complemented by a primary constraint
\be \beta = 0\label{beta}\ee
Commuting (calculating the Poisson brackets) the constraint with the Hamiltonian, we obtain  the secondary constraint
\be\label{c1} \{H,\beta\}= C_1 = \frac{1}{6}{\tilde{\alpha}^i}_j {\beta^j}_i - \frac{1}{3}p^{ac}g_{ac}=0\ee
In turn, commuting $C_1$ with the Hamiltonian, we obtain another secondary constraint
\be \{H, C_1\}=\frac{1}{12} \frac{{\beta^j}_i {\beta^i}_j }{\sqrt{-g}} - \frac{1}{2}\sqrt{-g}\left[{\rm tr} (\tilde\alpha^2)\right]^2 + {\tilde{\alpha}^i}_j p^{jk}g_{ki}=C_2\ee
Commuting this with the Hamiltonian no new constraints are generated.

Note that
\be H=-C_2+\alpha C_1\ee 
and thus the Hamiltonian vanishes on the constraint surface. This is natural in a conformal theory. Classically however, it only means that we should consider such solutions of equations of motion which have zero energy. The Hamiltonian is still an important quantity, as it generates the equations of motion, even though the energy vanishes on intersting classical trajectories.

Following the standard Dirac procedure, the first order constraint eq.(\ref{beta}) can be supplemented by another condition which turns the constraints into second order. A convenient choice is
\be \alpha=0\label{alpha}\ee
With this choice the Hamiltonian simplifies and we will adopt it in the following.

\subsection{General Linear Transformations}

Before analyzing equations of motion and their solutions, we note that our model has a large number of symmetries. We have already discussed gauge symmetry, which was inherited from the complete theory where original gauge transformations were taken to be independent of spatial coordinates.
However there is a larger subgroup of the original space-time dependent gauge group, which preserves the independence of the metric on $x_i$. These transformations appear in the reduced model not as gauge symmetries with associated constraints, but rather as global symmetries. The reason there are no constraints associated with these symmetries in the reduced model, is that they are automatically satisfied when the fields are taken to be $x_i$-independent.

Consider a general linear transformation that does not induce space dependence in the metric, and preserves the gauge conditions eq.(\ref{gaugef}). It's infinitesimal form is:

\be x'^\alpha = ({\delta^\alpha}_\beta + {\omega^\alpha}_\beta)x^\beta\ee
with ${\omega^0}_0 = 0; \ \ \ {\omega^0}_k=0$







 
 
 
 
 The transformation of the metric is:
  \begin{eqnarray} g_{ij}  &\rightarrow& g_{ij} - g_{ik} {\omega^k}_j - g_{kj}{\omega^k}_i\nonumber\\
  g^{ij}& \rightarrow& g^{ij} + g^{ik}{\omega^j}_k + g^{kj}{\omega^i}_k\end{eqnarray}
 For this to be a canonical transformation, the momenta have to transform as
 \be \delta p^{ij} = {\omega^i}_b p^{bj} +  {\omega^j}_b p^{ib}\ee 
 The transformation of $\alpha$ and $\beta$ can be found using the expression of $\alpha$ in terms of time derivative of $g$, and again requiring that the transformation is canonical
\begin{eqnarray}  {\alpha^i}_j &\rightarrow &{\alpha^i}_j (1+\frac{1}{3}\omega) - {\omega^k}_j {\alpha^i}_k + {\omega^i}_k {\alpha^k}_j\nonumber\\
{\beta^i}_j &\rightarrow& {\beta^i}_j (1-\frac{1}{3}\omega) - {\omega^k}_j {\beta^i}_k + {\omega^i}_k {\beta^k}_j\end{eqnarray}
 where, $\omega\equiv\omega^i_i $.
 
 It is indeed easy to check that this transformation leaves the Hamiltonian invariant. One has
 \be \delta H=\frac{\omega}{3} H\ee
 which vanishes on the constraint surface.
 
 The matrix $\omega_{ij}$ is an arbitrary real matrix, thus providing us with 9 symmetries. One of them, corresponding to $\omega_{ij}\propto \delta_{ij}$ however coincides with the conformal transformation. We should therefore strictly speaking consider only the traceless part $\omega_{ij}$ as generators of global symmetry transformations. The theory thus has 8 symmetries. With such large number of conserved quantities, as discussed in the introduction, one might hope that the dynamics of the model is stable. We will see however, that this is not the case. Nevertheless this large number of conserved quantity is handy to be able to find solutions of equations of motion.

\section{Solving the equations of motion}
Before directly tackling the solution of equations of motion it is useful to introduce a different set of coordinates, which simplifies this problem somewhat. At the moment our Hamiltonian is written in terms of basic variables $g_{ij}$ and $\tilde \alpha^i_j$. However not all of them are independent. The metric $g_{ij}$ is symmetric and contains 6 degrees of freedom, while $\tilde \alpha^i_j$ is not symmetric, but is nevertheless constrained sine $g_{ij}\tilde \alpha^j_k$ is by definition a symmetric matrix. Additionally, we set $\alpha=0$. Also the constraint eq.(\ref{c1}) can be used to eliminate one more degree of freedom. We can use it for example to fix $g=-1$. Thus in total we have 10 degrees of freedom. We will use the parametrization that makes these independent degrees of freedom more accessible.
 

We introduce the general real matrix $\Lambda$ by
\be\label{Lambda} g_{ij} = \ - \left[\Lambda \Lambda^T\right]_{ij}\ee
This relation defines $\Lambda$ only up to a rotation, as $\Lambda$ and $\Lambda O$ give the same matrix $g$.
To define it completely we take
\be\label{gamma}\tilde \alpha_{ij} = \left[{\Lambda^T}^{-1} \gamma \Lambda^T\right]_{ij}\ee
with $\gamma$ - a diagonal traceless matrix
\[ \gamma = \left| \begin{array}{ccc}
 \gamma_1 & 0 & 0 \\
0 & \gamma_2 & 0 \\
0 & 0 & -(\gamma_1 + \gamma_2) \end{array} \right|.\] 
With general $\gamma$ eq.(\ref{gamma}) is just a similarity transformation, but requiring $\gamma$ to be diagonal fixes the freedom in $\Lambda$ left undetermined by eq.(\ref{Lambda}). Tracelessnes of $\gamma$ follows from the tracelessnes of $\tilde\alpha$. The general matrix $\Lambda$ has 9 degrees of freedom, which we will reduce to 8 by requiring $\vert \Lambda\vert=1$. Together with two components of diagonal, traceless $\gamma$ this constitutes the original 10 degrees of freedom present in $\{g,\tilde\alpha\}$.

In terms of the new variables we have
\begin{eqnarray} \dot{g}&=& - (\dot{\Lambda} \Lambda^T + \Lambda \dot{\Lambda}^T)\\
\dot{\alpha}& =&  {\Lambda^T}^{-1} (\dot{\gamma} + \gamma {\dot{\Lambda}}^T {\Lambda^T}^{-1} -  {\dot{\Lambda}}^T {\Lambda^T}^{-1} \gamma  )= {\Lambda^T}^{-1} (D_0 \gamma) \Lambda^T\nonumber
\end{eqnarray}
where
\be D_0 \gamma \equiv \dot{\gamma} + [\gamma, M]; \ \ \ \ \ \ M \equiv {\dot{\Lambda}^T}{\Lambda^{T}}^{-1}\ee

The action eq.(\ref{action}) can now be written as :

\begin{align}
  \label{s}S &= -| \Lambda| \int dt \left\{3{\rm tr}\left(\dot{\gamma} ^2 + [\gamma, M]^2\right)  +\frac{1}{2} [{\rm tr}[\gamma^2]]^2 - {\rm tr}\ \tilde{\mu}\left[\gamma  - \left(M+M^T  \right)\right] \right. \nonumber\\
 &\qquad \left. + \alpha {\rm tr}[\gamma \dot{\gamma} ]+ \frac{1}{3}\mu[\alpha - 2{\rm tr}M] +\frac{1}{2}\alpha^2 {\rm tr}[\gamma^2] \right\}
\end{align}

The Lagrange multiplier (symmetric) matrix $\tilde\mu$ enforces the constraint relating $\tilde\alpha$ to time derivative of $g$.
Just like in the previous section, we can set $\alpha=0$, since there is no time derivative of $\alpha$ in eq.(\ref{s}). This can be done, but only after requiring that the variation of $S$ with respect to $\alpha$ vanishes. This variation leads to a constraint
\be \frac{\partial S}{ \partial \alpha} \Bigr{ |}_{ \alpha=0}=|\Lambda|({\rm tr}\gamma \dot{\gamma} + \frac{1}{3}\mu)=0; \label{alpha0}\ee
This is the generator of the conformal gauge transformation expressed in the new variables.

Calculating momenta conjugate to $\Lambda$, we find
\be \label{p}p_{ij} = \frac{\partial L}{\partial \dot{\Lambda}_{ij}}= -  |\Lambda| \left[ {\Lambda^T}^{-1} \left(6 [[\gamma, M],\gamma] + 2 (\tilde{\mu} - \frac{1}{3}I \mu)\right)\right]_{ij}\ee

 Note that on the constraint surface the symmetric part of matrix $M$ is proportional to $\gamma$. Thus only the antisymmetric part of $M$ contributes to the commutator in eqs.(\ref{s},\ref{p}). Using this, we find
\begin{eqnarray} \frac{1}{2}(\Lambda^T p-p\Lambda^T) &=& -6  |\Lambda| [[\gamma,\frac{1}{2}(M-M^T)],\gamma] \label{aa}\nonumber\\
 \frac{1}{2}(\Lambda^T p +p\Lambda^T)&=& -2  |\Lambda|(\tilde{\mu}- \frac{1}{3}I\mu)
 \end{eqnarray} 
 
Conjugates to $\gamma$ are found as

\be p_1 =\frac{\partial L}{\partial \dot{\gamma}_1}= -  6 |\Lambda|  (2 \dot{\gamma_1} + \dot{\gamma_2}), \ \ \ \ \ \ p_2 = \frac{\partial L}{\partial \dot{\gamma}_2}=-   6 |\Lambda| (2 \dot{\gamma_2} + \dot{\gamma_1}) \ee

The Hamiltonian is: 
\be H=\frac{1}{2} \Lambda^T p \gamma - 3 |\Lambda|[\gamma, \frac{1}{2}(M-M^T)]^2 + \frac{1}{18 |\Lambda|}(-p_1^2 - p_2^2 + p_1 p_2) + \vert\Lambda\vert(\gamma_1^2 + \gamma_2^2 + \gamma_1 \gamma_2)^2 \ee
It is now possible to express the second term in terms of conjugate momenta using eq.(\ref{aa}). It is most simply done by expanding both sides of eq.(\ref{aa}) in terms of the complete basis of $3\times 3$ matrices.
After some straightforward algebra, we find:

\be [\gamma, \frac{1}{2}(M-M^T]^2= \frac{1}{18 |\Lambda|^2} \left[\left(\frac{(\Lambda^T p - p \Lambda^T)_{12}}{\gamma_2 - \gamma_1}\right)^2 + \left(\frac{(\Lambda^T p - p \Lambda^T)_{13}}{\gamma_2 + 2 \gamma_1}\right)^2 + \left(\frac{(\Lambda^T p - p \Lambda^T)_{23}}{2\gamma_2 + \gamma_1}\right)^2\right]\ee
Finally, diagonalizing the quadratic term in the Hamiltonian, we obtain: 

\begin{eqnarray} H= &-&\frac{1}{18|\Lambda|}[\tilde{p_1}^2 + \tilde{p_2}^2]+ \frac{9}{16} \vert\Lambda\vert[\tilde{\gamma}_1^2 + \tilde{\gamma}_2^2]^2  + \frac{1}{2} {\rm tr}\left(\Lambda^T p \gamma\right)\nonumber\\
& -&  \frac{1}{6 |\Lambda|} \left[\left(\frac{(\Lambda^T p - p \Lambda^T)_{12}}{\gamma_2 - \gamma_1}\right)^2 + \left(\frac{(\Lambda^T p - p \Lambda^T)_{13}}{\gamma_2 + 2 \gamma_1}\right)^2 + \left(\frac{(\Lambda^T p - p \Lambda^T)_{23}}{2\gamma_2 + \gamma_1}\right)^2\right] \end{eqnarray}

Where, 
\be \tilde{p_1} = \frac{1}{2}(p_1 + p_2), \ \ \ \ \ \tilde{p_2} = \frac{\sqrt{3}}{2} (-p_1 + p_2)\ee
and
\be  \tilde{\gamma_1} = (\gamma_1 + \gamma_2), \mbox{   } \tilde{\gamma_2} = \frac{1}{\sqrt{3}} (-\gamma_1 + \gamma_2)\ee

The canonical form of the constraint eq.(\ref{alpha0}), which supplements this Hamiltonian is:
\be \frac{1}{3}(p_1\gamma_1 + p_2 \gamma_2) + {\rm tr}(\Lambda^T p)=0\ee
As noted above, we fix the gauge freedom associated with this constraint by setting $|\Lambda|=1$\cite{footnote2}.

Our goal here is to see whether the Hamiltonian has unstable solutions. We will not look for a general solution of equations of motion, but instead will analyze a simple subset of those. The simplification is possible due to the following observation. Let us define for convenience traceless matrices
\be\tau_1={\rm diag} (1,0,-1); \ \ \ \tau_2={\rm diag} (0,1,-1); \ \ \ \sigma^{a}_{ij}=\epsilon_{aij}\ee
\be \lambda^1 = \left| \begin{array}{ccc}
 0& 0 & 0 \\
0& 0 & 1 \\
0 & 1 &0 \end{array} \right| ; \   \lambda^2 = \left| \begin{array}{ccc}
 0& 0 & 1 \\
0& 0 & 0 \\
1 & 0 &0 \end{array} \right| ; \   \lambda^3 = \left| \begin{array}{ccc}
 0& 1 & 0 \\
1& 0 & 0 \\
0 & 0 &0 \end{array} \right|\ee
and associated generators of the general linear transformations
\be G^i={\rm tr}(\Lambda^Tp\tau^i); \ \ \ G^a_A={\rm tr}(\Lambda^Tp\sigma^a); \ \ \ G^a_S={\rm tr}(\Lambda^Tp\lambda^a)\ee
In terms of these, the Hamiltonian is written 

\be \label{ha}H= -\frac{1}{18|\Lambda|}[\tilde{p_1}^2 + \tilde{p_2}^2]+ \frac{9}{16} \vert\Lambda\vert[\tilde{\gamma}_1^2 + \tilde{\gamma}_2^2]^2  + \frac{1}{2} \Sigma_i\left(G^i\gamma_i\right) -  \frac{1}{6 |\Lambda|} \left[\left(\frac{G^3_A}{\gamma_2 - \gamma_1}\right)^2 + \left(\frac{G^2_A}{\gamma_2 + 2 \gamma_1}\right)^2 + \left(\frac{G^1_A}{2\gamma_2 + \gamma_1}\right)^2\right] \ee

Note that for all of these generators, we have $[|\Lambda|,G^\alpha]=0$.
Consider a solution, which at initial time has $G^i=G^a_A=G^a_S=0$. Since commutator of any of the generators $G^\alpha$ with the Hamiltonian eq.(\ref{ha}) is proportional to, at least  the first power of $G^\beta$, this condition is preserved in time, and all the generators $G^\alpha$ vanish at all times. 
We can think of this initial condition, as an initial condition imposed on $p_{ij}$ for arbitrary initial $\Lambda_{ij}$. For this set of initial conditions, the equations of motion therefore simplify considerably. The equation of motion for $\Lambda$ becomes
\be \dot{\Lambda}_{ij} = \frac{1}{2}(\gamma\Lambda)_{ij} \ee 
This determines $\Lambda$ once the solution for $\gamma$ is known as
\be \Lambda = A\left( \exp{\int^t_0\frac{\gamma}{2}dt}\right)\ee
where $A$ is the initial condition.

The equations of motion for $\gamma$ then are derived from he reduced Hamiltonian
\be H= -\frac{1}{18}[\tilde{p_1}^2 + \tilde{p_2}^2]+ \frac{9}{16} [\tilde{\gamma}_1^2 + \tilde{\gamma}_2^2]^2 \ee 
where we have set $|\Lambda|=1$ with accordance to previous discussion.

The reduced Hamiltonian is a simple upside-down unharmonic oscillator. The kinetic term is negative, in accordance with the fact that $\gamma_i$ appear as ghost modes in the linearized theory, where the unharmonic potential is absent. Interestingly, the sign of the potential is positive, and therefore it is clear that the dynamics of the reduced model is unstable. To see this explicitly, consider a simple solution of equations of motion, corresponding to vanishing "`angular momentum" in the $\tilde\gamma_1-\tilde\gamma_2$ plane. We also have to impose the constraint of zero energy, which is an easy task in the reduced model. Solutions under these conditions are very simple
\be \tilde\gamma_1=\gamma_r\ \cos\theta ; \ \  \tilde\gamma_2=\gamma_r\ \sin\theta\ee
with
\be\label{solution}\theta=const; \ \ \gamma_r = \frac{\gamma_0}{1 \pm \frac{\gamma_0}{2^{3/2}} t}\ee
The two solutions correspond to the sign of the initial radial velocity. For negative initial velocity (sign $+$ in eq.(\ref{solution})), the "`particle" initially moves towards the origin. This is a stable solution, since at infinite time the particle simply climbs to the top of the potential, and ends up there with zero velocity. For positive initial relative velocity (sign $-$ in eq.(\ref{solution}))
the particle moves away from the origin. This solution is unstable. The instability is in fact much worse than would be for an upside down harmonic oscillator. The particle reaches infinite distance within a finite time $t_c=2^{3/2}/\gamma_0$.

Transforming to the original variables we find
\be \gamma_{1,2} = \frac{1}{2}(\cos{\theta} \mp \sqrt{3}\sin{\theta}) \gamma_r=\frac{1}{2}(\cos{\theta} \mp \sqrt{3}\sin{\theta}) \frac{\gamma_0}{1 \pm \frac{\gamma_0}{2^{3/2}} t}\ee
The metric $g$ is found to be
\be\label{cosm} g_{ij}=-[A\Gamma A^T]_{ij}\ee
where $\Gamma$ is the diagonal matrix with the following non-vanishing matrix elements
\be \label{cosm1}\Gamma_{11}=|1 \pm \frac{\gamma_0}{2^{3/2}} t|^{2^{3/2}\left(\cos{\theta} - \sqrt{3}\sin{\theta}\right)}; \ \ \ \Gamma_{22}=|1 \pm \frac{\gamma_0}{2^{3/2}} t|^{2^{3/2}\left(\cos{\theta} + \sqrt{3}\sin{\theta}\right)};\ \ \Gamma_{33}=[\Gamma_{11}\Gamma_{22}]^{-1}\ee

Either one or two eigenvalues of the metric $g$ diverge at the terminal time $t_c$, while the rest of the eigenvalues (two or one) vanish.

\section{Discussion.}
In this paper we have considered conformal gravity in translationally invariant approximation. Our main finding is, that the nonlinear interactions lead to instability in the dynamics of zero momentum modes. Specifically we displayed a simple solution of equations of motion which diverges within a finite time. The reason for such a severe divergence is that the dynamical modes $\gamma$, which in the perturbative regime have ghostlike kinetic term, acquire in addition a positive potential. Thus this sector of the reduced theory is equivalent to two dimensional upside down anharmonic oscillator. 
Close to the minimum of the potential $\gamma$ behaves as a pertubative ghost with zero mass. However at any non-vanishing distance from the minimum, the signs of kinetic and potential energies are opposite and $\gamma$ behaves as a tachyon. 

Thus the perturbative ghost problem is not cured, but is rather exacerbated by nonlinear gravitational interactions. Thinking about quantization, it is clear that the theory does not allow sensible quantization via standard methods, i.e. using standard Dirac norm. The possibility that the use of a nonstandard norm, like in \cite{mann} could lead to a unitary theory may be worth exploring, although such a procedure is rather non intuitive.

Finally we note that another way to view the present calculation is as a study of possible homogeneous cosmologies in conformal gravity. The universe described by eqs.(\ref{cosm},\ref{cosm1}) is certainly very far from reality, since it is not isotropic. In fact the only isotropic and homogeneous space allowed by conformal gauge symmetry is Minkowski space, since any isotropic metric is conformally equivalent to Minkowski one. Nevertheless, an interesting property of this metric, is that it describes accelerated dynamics. As we indicated above, some dimensions in this space undergo accelerated expansion, while others accelerated contraction. Perhaps, when supplemented by conformal anomaly in the matter part\cite{thooft}, which we have not considered here, it could acquire more realistic features while still retaining the property of acceleration. This would be interesting to study.

\section{Appendix: Residual gauge symmetry of the action}
In this appendix we show that the action eq.(\ref{action}) after gauge fixing is still invariant under a combination of a general linear and conformal transformation which has not been gauge fixed by eq.(\ref{gaugef}).

Under a combined transformation the metric transforms as
\be g_{\rho\sigma} (x)\rightarrow g_{\rho\sigma}' (x') = \Omega^2 (x)g_{\mu\nu}(x) \frac{\partial x^\mu}{\partial {x^\rho}'}\frac{\partial x^\nu}{\partial {x^\sigma}'}\ee
In order for the metric to remain a function of time only, we must only consider the transformation of the type
\be x^i={x^i}',\ \ \  x^0=f({x^0}'),\ \ \ \Omega=\Omega(t)\ee
With this restriction we get $g_{i0}'(x') = 0$ if $g_{i0}(x)=0$, thus this gauge fixing condition is preserved. In order to maintain the condition $g_{00}(x')=1$, we need to take $\Omega^2(t)=\frac{1}{f'^2}$.
The spatial components of the metric transform under this transformation as
\be g_{ij}(t)\rightarrow g'_{ij}(t')=\frac{1}{f'^2}g_{ij}(t(t'))\ee. 
Denoting $\frac{1}{f'}=F$, we can write
\be g'_{ij}(t)= F^2 g_{ij}(f(t)), \ \ \ \ \ g'^{ij}(t)=\frac{1}{F^2}g^{ij}(f(t))\ee
Then, using
\be \frac{\partial}{\partial t} = \frac{1}{F}\frac{\partial}{\partial f}\ee
we obtain
\be \partial_t g_{ij}(t)\rightarrow \partial_t g'_{ij}(t)= \partial_t {F^2}g_{ij}(f) + F^2 \partial_t g_{ij}(f)= \partial_t (F^2)g_{ij}+ F \partial_f g_{ij}\ee
and
\be {\alpha^k}_i(t)\rightarrow{\alpha'^k}_i(t)=g'^{kj}\partial_t g'{ij} = \frac{1}{F^2}g^{kj}[\partial_t{F^2}g_{ij}+ F \partial_f g_{ij}]= \frac{\partial_t{F^2}}{F^2}{\delta^k}_i + \frac{1}{F}{\alpha^k}_i(f)\ee,
Or
\be {{\tilde{\alpha}}^k}_j(t)\rightarrow \frac{1}{F}{{\tilde{\alpha}^k}}_j(f); \ \ \ \ \  \alpha(t) \rightarrow 3\frac{\partial_t (F^2)}{F^2} + \frac{1}{F}\alpha(f)\ee

Similarly, it follows that:

\be \partial_t {\alpha^k}_j(t)\rightarrow\partial_t {\alpha'^k}_j(t)=\partial_t(\frac{\partial_t(F^2)}{F^2})\delta^k_j + \partial_t (\frac{1}{F}){\alpha^k}_j(f) + \frac{1}{F^2}\partial_f {\alpha^k}_j(f)\ee
Or
\be \partial {{\tilde{\alpha}}^k}_j(t) \rightarrow \partial_t (\frac{1}{F}){{\tilde{\alpha}}^k}_j(f) + \frac{1}{F^2}\partial_f {{\tilde{\alpha}}^k}_j(f); \ \ \ \ \ \partial_t\alpha(t) \rightarrow 3\partial_t (\frac{\partial_t (F^2)}{F^2}) + \partial_t (\frac{1}{F})\alpha(f) + \frac{1}{F^2}\partial_f \alpha(f)\ee
It is now straightforward to substitute these transformed fields in the expression for the action eq.(\ref{action}). Upon discarding total derivative terms and changing the integration variables $t\rightarrow f$ it is then easy to see that the action is indeed invariant.

\section*{Acknowledgments}
The research  was supported by the DOE grant DE-FG02-13ER41989.


\begin{thebibliography}{99}
\bibitem{footnote1}We note that it may be possible to make sense of conformal gravity even if ghosts do cause instability by either restricting oneself to a subset of solutions \cite{malda}, or invoking quantization with nonstandard inner product \cite{mann}. However we are not aware at the moment of a fully consistent way of implementing either of these suggestions in conformal gravity.
\bibitem{footnote2} We do this only after deriving equations of motion to avoid the necessity to introduce Dirac brackets.
\end{thebibliography}
\end{document}